\documentclass[12pt]{article} 
\pdfoutput=1

\usepackage[utf8]{inputenc}
\usepackage[T1]{fontenc} 
\usepackage{lmodern}
\usepackage[english]{babel}           
\usepackage{cancel}
\usepackage[pdftex]{graphicx}
\usepackage{epsfig}
\usepackage{graphicx}
\usepackage{comment}
\usepackage{latexsym}
\usepackage{hyperref}
\usepackage{amsmath}
\usepackage{mathrsfs}
\usepackage[usenames, dvipsnames]{color}
\usepackage{amsbsy}
\usepackage{amssymb}
\usepackage{amsthm}
\usepackage{amsfonts}
\usepackage{cite}

\usepackage{xcolor}
\usepackage{color}
\usepackage{mathtools}
\usepackage{caption} 
\usepackage{subcaption} 
\usepackage{euscript} 
\usepackage{float}
\usepackage{diagbox}
\usepackage[title]{appendix}
\usepackage{tabularx}

\newcommand{\N}{{\cal N}}
\newcommand{\T}{{\cal T}}
\newcommand{\A}{{\cal A}}
\renewcommand{\H}{{\cal H}}
\newcommand{\F}{{\cal F}}
\newcommand{\D}{{\cal D}}
\newcommand{\J}{{\cal J}}
\renewcommand{\i}{{\rm i}}
\newcommand{\f}{{\rm f}}
\newcommand\q{{\rm q}}

\newcommand\Vol{{\rm Vol}}
\newcommand\Diff{{\rm Diff}}

\renewcommand\d{{\rm d}}

\DeclareMathOperator\sign{sign}

\DeclareMathOperator\Rep{Re}
\DeclareMathOperator\Imp{Im}

\newcommand{\be}{\begin{equation}}
\newcommand{\ee}{\end{equation}}
\newcommand{\dis}{\displaystyle}
\renewcommand{\thefootnote}{\fnsymbol{footnote}}

\newcommand{\Eq}[1]{Eq.~\eqref{#1}}
\newcommand{\Eqs}[1]{Eqs.~\eqref{#1}}
\newcommand{\Refe}[1]{Ref.~\cite{#1}}
\newcommand{\Refs}[1]{Refs.~\cite{#1}}
\newcommand{\Sect}[1]{Sect.~\ref{#1}}

\newcommand{\R}{\mathbb{R}}

\renewcommand{\O}{{\cal O}}
\newcommand{\cst}{{\rm cst.}}

\newcommand{\ie}{{\em i.e.} }
\newcommand{\eg}{{\em e.g.} }

\newcommand{\apriori}{{\it a priori} }
\newcommand{\where}{\mbox{where}}

\renewcommand{\and}{\mbox{and}}

\newcommand{\esps}{\phantom{\!\!\!\overset{|}{a}}}
\newcommand{\esp}{\phantom{\!\!\overset{\displaystyle |}{|}}}

\newcommand{\bm}{\boldmath} 

\newcommand{\black}{\color{black}}

\catcode`\@=11
\def\marginnote#1{}
\newcount\hour
\newcount\minute
\newtoks\amorpm
\hour=\time\divide\hour by60 \minute=\time{\multiply\hour by60
\global\advance\minute by-\hour}
\edef\standardtime{{\ifnum\hour<12 \global\amorpm={am}%
        \else\global\amorpm={pm}\advance\hour by-12 \fi
        \ifnum\hour=0 \hour=12 \fi
        \number\hour:\ifnum\minute<10 0\fi\number\minute\the\amorpm}}
\edef\militarytime{\number\hour:\ifnum\minute<10 0\fi\number\minute}
\def\draftlabel#1{{\@bsphack\if@filesw {\let\thepage\relax
   \xdef\@gtempa{\write\@auxout{\string
      \newlabel{#1}{{\@currentlabel}{\thepage}}}}}\@gtempa
   \if@nobreak \ifvmode\nobreak\fi\fi\fi\@esphack}
        \gdef\@eqnlabel{#1}}
\def\@eqnlabel{}
\def\@vacuum{}
\def\draftmarginnote#1{\marginpar{\raggedright\scriptsize\tt#1}}
\def\draft{\oddsidemargin -.2truein
        \def\@oddfoot{\sl preliminary draft \hfil
        \rm\thepage\hfil\sl\today\quad\militarytime}
        \let\@evenfoot\@oddfoot \overfullrule 3pt
        \let\label=\draftlabel
        \let\marginnote=\draftmarginnote
   \def\@eqnnum{(\theequation)\rlap{\kern\marginparsep\tt\@eqnlabel}%
\global\let\@eqnlabel\@vacuum}  }
\def\thebibliography#1{
\vskip 0.5cm \centerline{\bf \Large References}
\list{
[\arabic{enumi}]}{\settowidth\labelwidth{[#1]}
\leftmargin\labelwidth
\advance\leftmargin\labelsep
\usecounter{enumi}}
\def\newblock{\hskip .11em plus .33em minus .07em}
\sloppy\clubpenalty4000\widowpenalty4000
\sfcode`\.=1000\relax}

\renewcommand{\theequation}{\arabic{section}.\arabic{equation}}
\renewcommand{\section}{\setcounter{equation}{0}\@startsection
{section}{1}{0mm}{-\baselineskip}{0.5\baselineskip} {\normalfont\Large\bfseries}}
\renewcommand{\subsection}{\@startsection
{subsection}{2}{0mm}{-\baselineskip}{0.5\baselineskip} {\normalfont\large\bfseries}}
\renewcommand{\subsubsection}{\@startsection
{subsubsection}{3}{0mm}{-\baselineskip}{0.5\baselineskip}
{\normalfont\normalsize\slshape}}

\begin{document}


\begin{titlepage}
\begin{flushright}
CPHT-RR081.112024, December 2024
\vspace{0.0cm}
\end{flushright}
\vspace{6mm}

\begin{centering}
{\bm\bf \Large  The exact Wheeler-DeWitt equation for the \\ \vspace{.2cm} scale-factor minisuperspace model}

\vspace{9mm}

 {\bf Eftychios Kaimakkamis,$^1$\footnote{kaimakkamis.eftychios@ucy.ac.cy} Herv\'e Partouche,$^2$\footnote{herve.partouche@polytechnique.edu} Karunava Sil$^{1, 3}$\footnote{karunavasil@gmail.com} \\ \vspace{0.1cm} and Nicolaos Toumbas$^1$\footnote{nick@ucy.ac.cy}}

 \vspace{4mm}

$^1$ {\em Department of Physics, University of Cyprus, \\Nicosia 1678, Cyprus}

$^2$  {\em CPHT, CNRS, Ecole polytechnique, IP Paris, \\F-91128 Palaiseau, France}

$^3${\em Department of Physics, New Alipore College,\\L Block, New Alipore, Kolkata 700053, India}

\end{centering}
\vspace{0.2cm}
$~$\\
\centerline{\bf\Large Abstract}\\
\vspace{-0.8cm}

\begin{quote}

We consider the classical minisuperspace model describing a closed, homogeneous and isotropic Universe, with a positive cosmological constant. Upon canonical quantization, the infinite number of possible operator orderings in the quantum Hamiltonian leads to distinct Wheeler-DeWitt equations for the wavefunction $\psi$ of the Universe. Similarly, ambiguity arises in the path-integral formulation of~$\psi$, due to the infinite choices of path-integral measures for the single degree of freedom of the model. For each choice of path-integral measure, we determine the correct operator ordering of the quantum Hamiltonian and derive the exact form of the Wheeler-DeWitt equation, including all corrections in $\hbar$. Additionally, we impose Hermiticity of the Hamiltonian to deduce the Hilbert-space measure $\mu$, which appears in the definition of the Hermitian inner product. Remarkably, all quantum predictions can be expressed through the combination $\Psi=\sqrt{\mu}\psi$, which satisfies a universal Wheeler-DeWitt equation, independent of the initial path-integral prescription. This result demonstrates that all seemingly distinct quantum theories arising from different quantization schemes of the same classical model are fundamentally equivalent.

\end{quote}


\end{titlepage}
\newpage
\setcounter{footnote}{0}
\renewcommand{\thefootnote}{\arabic{footnote}}
 \setlength{\baselineskip}{0.7cm} \setlength{\parskip}{.2cm}

\setcounter{section}{0}


\section{Introduction}

There are a number of unresolved questions concerning the Wheeler-DeWitt  (WDW) equation for the wavefunction of the Universe \cite{DeWitt:1967yk}, both conceptual and technical \cite{DeWitt:1967yk, Vilenkin:1982de, Vilenkin:1983xq, Vilenkin:1984wp, Vilenkin:1987kf,  Hartle:1983ai, Hawking:1984hk, Halliwell:1984eu, Hawking:1985bk, Halliwell, Halli2, Hartle:2008ng, Turok1, Halli-Hartle, Linde:1990flp}. Any progress towards their resolution will be important for our understanding of quantum cosmology at a fundamental level. In this work we will focus on a technical question regarding the existence of operator-ordering ambiguities in implementing the Hamiltonian constraint at the quantum level---see \eg \cite{DeWitt:1967yk,Hawking:1985bk, Halliwell}. Despite the fact that a number of interesting proposals have been put forward over the years to resolve this issue, imposing for example special boundary conditions for the wavefunction of the Universe \cite{Nelson:2008vz, He:2015wla}, a complete satisfactory solution is still lacking. 

Indeed, even in the simplest case of a minisuperspace model involving a single degree of freedom, there is an infinite number of operator-ordering choices in the expression of the quantum Hamiltonian, leading to distinct WDW equations for the wavefunction of the Universe. It does not seem possible to select among these choices from first principles. The same ambiguities arise in the path-integral formulation of the quantum theory. Indeed, while the classical action is invariant under field redefinitions of the minisuperspace degree of freedom---here the scale-factor degree of freedom---the gauge-invariant path-integral measure is not, since the different  measures are related by no-trivial Jacobians. There is indeed an infinite number of inequivalent gauge-invariant path-integral measures, leading to different expressions for the wavefunction of the Universe and distinct WDW equations \cite{PTV}. Naively, it seems that the same classical theory leads to inequivalent quantum theories. It is important to stress, however, that in order for the quantum prescriptions to be truly inequivalent, they must lead to distinct predictions for the physical observables and probability amplitudes. To address the question fully, we must also obtain an appropriate inner product on the Hilbert space of wavefunctions, which can be utilized to define probability amplitudes and other physical observables.  

In previous work \cite{PTV, Partouche:2022kfi, Partouche:2021lyb},  two of us studied the issue of operator-ordering ambiguities in the simplest one-dimensional minisuperspace model arising in the context of pure four-dimensional Einstein gravity in the presence of a positive cosmological constant. In this approximation, we take the cosmological metric to be homogeneous and isotropic, and the spatial sections to be closed. There is a single degree of freedom, namely the scale factor of the cosmology.  Our analysis in that work focussed on the semiclassical limit. For each path-integral wavefunction based on a gauge-invariant measure associated to a field redefinition of the scale factor, we determined the ambiguity functions and the WDW equation it satisfies {\it at the semiclassical level}. The inner-product measure on the Hilbert space was determined by requiring the Hamiltonian operator to be Hermitian. In fact, this Hermiticity condition suffices to determine the inner-product measure uniquely for each such quantum prescription. Taking the inner-product measure into account, we demonstrated that the various quantum prescriptions lead to identical results for the probability amplitudes at the semiclassical level~\cite{PTV, Kehagias:2021wwr}. This result was taken to be suggestive that the various quantum prescriptions are in fact equivalent, with the universality extending beyond the semiclassical level.

However, our analysis in \Refe{PTV} did not determine all the ambiguity functions that can appear in the WDW equation, since the computations were restricted at the semiclassical level. Additional ambiguity can arise at higher orders in the $\hbar$ expansion, which can be resolved by computing the path-integral wavefunction at higher orders. It is important to investigate whether the universality of the physical observables persists at the exact order in $\hbar$, rendering the multitude of the quantum prescriptions associated with the path-integral formulation physically equivalent. In the present work, we show that this is indeed the case. By discretizing suitably the path-integral wavefunction for each quantum prescription, and taking the continuum limit~\cite{Peskin}, we obtain the WDW equation it satisfies exactly, to all orders in $\hbar$, and, therefore, resolve all ambiguity functions in the expressions. Physical observables and probability amplitudes are defined by utilizing the Hermitian inner-product measure, which we also determine exactly. We demonstrate that the various quantum prescriptions yield identical results for the physical observables to all orders in $\hbar$. In fact, the product of the path-integral wavefunction with the square root of the inner-product measure satisfies a universal WDW equation, which is free of any ambiguities associated with operator-ordering choices or the path-integral measure. This result establishes the universality of the various quantum prescriptions at the exact level in $\hbar$. 

The plan of the paper is as follows. In Sect.~\ref{classmod}, we review the classical one-dimensional minisuperspace model based on four-dimensional pure Einstein gravity in the presence of a positive cosmological constant. We cast it as a one-dimensional non-linear $\sigma$-model involving a one-dimensional target space. In Sect.~\ref{quanti}, we carry the canonical quantization of the model. We show how to parametrize the operator-ordering ambiguities in the Hamiltonian and the WDW equation. This involves two complex functions of the minisuperspace degree of freedom, transforming as a scalar and vector under field redefinitions, respectively. In Sect.~\ref{path}, we show how the path-integral approach to quantization also leads to ambiguities due to the multitude of choices for the gauge-invariant path-integral measure, associated with field redefinitions. In Sect.~\ref{inner pro}, we impose the Hermiticity condition of the quantum Hamiltonian in order to obtain the inner-product measure for each quantum prescription or path-integral measure choice, and use it to define probability amplitudes. The latter depend on the combination $\Psi=\sqrt{\mu}\psi$, where $\mu$ is the inner-product measure and $\psi$ is the wavefunction of the Universe. 
We proceed in Sect.~\ref{resolving} to derive the ambiguity functions and the exact WDW equation, valid to all orders in the $\hbar$ expansion, for each choice of path-integral measure. This is achieved by considering a suitable discretized version of the path-integral wavefunction and taking the continuum limit at the end. Having obtained the ambiguity functions, we show that for all these quantum prescriptions, $\Psi$ satisfies a universal equation, to all orders in $\hbar$. Therefore, the probability amplitudes and physical observables are universal. In Sect.~\ref{compa}, we comment on previous results appearing in the literature. Sect.~\ref{conclu} explains why our approach is limited to the case where the $\sigma$-model target space is locally Minikowskian, when applied to minisuperspace models involving more than one degree of freedom. We also conclude with future research directions.


\section{Classical minisuperspace model}
\label{classmod}

In this section, we present the simplest classical minisuperspace model in four dimensions, which will  be quantized in the following sections.

Let us consider four-dimensional cosmologies, in the presence of a positive cosmological constant $\Lambda$ and in the absence of matter.\footnote{The four-dimensional pure Einstein theory with vanishing cosmological constant does not yield non-trivial homogeneous and isotropic cosmological solutions with  compact closed spatial sections and constant positive curvature. Non-isotropic homogeneous solutions with flat toroidal spatial sections do exist, but to realize these we need to retain more than one metric degree of freedom in the minisuperspace analysis. In this work, we will analyze only one-dimensional minisuperspace models.} In General Relativity, this can be described from the Lorentzian action 
\begin{equation}\label{action1}
S= \frac{1}{2}\int \limits_{ \mathcal{M}} { \d}^4x \sqrt{-g}\left(R-2\Lambda\right){ -} \int \limits_{\partial \mathcal{M}}\d^3y \sqrt{h} \, K\,,
\end{equation}
where $\mathcal{M}$ denotes the spacetime manifold of metric $g_{\mu\nu}$ and Ricci curvature $R$.\footnote{We use units where the Newton constant satisfies $8\pi G=1$. } The second term is the Gibbons-York-Hawking boundary term, which contributes when the manifold~$\mathcal{M}$ has a boundary~$\partial \mathcal{M}$. In the present case, $\partial \mathcal{M}$ is spacelike and we denote its coordinates as~$y^i$, induced metric as~$h_{ij}$ and trace of the extrinsic curvature as~$K$. 

For a closed, homogeneous and isotropic Universe, space is a 3-sphere and the metric takes the following {Friedmann-Lema\^itre-Robertson-Walker form\footnote{It is essential for our formalism to consider spatially compact cosmological models, for which the minisuperspace Lagrangian and action are finite, and the formulation of the wavefunction of the universe in terms of a path integral is possible. If the spatial sections were non-compact, the minisuperspace action would be infinite, as this is proportional to the volume of the spatial sections, and the wavefunction path integral could not be defined. Flat or open homogeneous and isotropic universes with non-compact spatial sections of vanishing and constant negative curvature, respectively, cannot be considered. The open cases (for which the spatial sections have constant negative curvature) would require non-trivial compactifications and boundary conditions, breaking isotropy and possibly homogeneity. So we do not consider these in this work that focusses on simple one-dimensional minisuperspace models.  Also, when the cosmological constant is vanishing, non-trivial homogeneous flat four-dimensional cosmological solutions, with compact toroidal sections,  cannot be realized in the context of simple one-dimensional minisuperspace models. For positive cosmological constant, we could consider the case of a homogeneous model with a single minisuperspace degree of freedom (single scale factor degree of freedom) and flat toroidal spatial sections, breaking global isotropy. In this case, all of our conclusions and methods concerning the ordering ambiguities presented in the next section apply since, as we will argue, the kinetic term in the Hamiltonian remains unaffected. }  
\begin{equation}
\label{FRW}
\d s^{2} = -N^{2}(x^0) {\d x^{0}}^2 + a^2(x^0)\d\Omega_{3}^2\,.
\end{equation}
In this  expression, $N(x^0)$ is the lapse function, $a(x^0)$ is the scale factor and $\d\Omega _{3}^{2}$ is the metric on the unit 3-sphere. The lapse function is non-dynamical and can be fixed to unity by a suitable time reparametrization, while $a(x^0)$ is a scalar under time reparametrizations. Inserting the above metric in the action, we obtain the effective Lagrangian for the scale-factor degree of freedom,
\begin{equation} \label{lagrangian}
L= 3v_{3}N \left(-\frac{a}{N^2}\, \dot{a}^{2}+a-\lambda ^{2}a^3\right),\qquad \where \qquad \lambda \black = \sqrt{\Lambda\over 3}\,.
\end{equation}
In this equation, $v_{3} = 2\pi^{2}$ is the volume of the unit $3$-sphere and dots stand for derivatives with respect to time $x^0$. This Lagrangian is that of  the minisuperspace model, where all inhomogeneities and the other metric degrees of freedom are omitted. Varying the action with respect to $N$ leads to an equation for the scale factor. Using cosmological time $t$, which corresponds to the gauge choice $N(t)\equiv 1$, this equation becomes
\begin{equation} \label{friedmann1}
 (\lambda a)^2 -\left({\d(\lambda a)\over \d(\lambda t)}\right)^2=1\,.
\end{equation}
Its solution consists of a pure de Sitter space, namely
\begin{equation} \label{scalefactor}
    \lambda a(t) = \cosh(\lambda t)\, .
\end{equation}
Note that this one-dimensional minisuperspace model does not contain any propagating physical degrees of freedom, and the full solution at the classical level can be obtained by solving a first-order constraint equation (obtained after fixing time-reparametrization invariance). The equation of motion for the scale factor follows by differentiating this constraint equation. See Ref.~\cite{Henneaux:1992ig} for further discussions.

We may consider field redefinitions of the scale factor, 
\be
\label{aA}
a=\A(q)\,,
\ee 
where $\A$ is an arbitrary positive, invertible function of a new field $q$. Such field redefinitions of the scale factor may prove useful and practical in finding solutions of the WDW equation at the quantum level. In addition, they are linked with ambiguities in the path-integral quantization of the model, as we explain explicitly in \Sect{path}. In terms of  $q(x^0)$, the classical Lagrangian in \Eq{lagrangian} becomes 
\begin{equation}
 {L}= 3v_{3}N \left(-\frac{\A\A'^2}{N^2}\,\dot q^2+\A-\lambda ^{2}\A^3\right),
 \label{Lag}
\end{equation}
where primes denote  derivatives with respect to $q$.
It is useful to interpret the model as a non-linear {$\sigma$-model,} with the base manifold---or domain---being a line parametrized by time and the target space being a one-dimensional manifold $\cal T$ parametrized by the values of the field  $q$. In order to avoid later confusions in the notations, we will denote the corresponding coordinate in $\T$ in roman style, $\q$. In this case, the Lagrangian can be written in the form
\begin{equation} \label{minisuperspace}
    L(N,q,\dot q) = N \left({1\over 2}\, \gamma _{\q\q}(q)\, \frac{\dot{q}^{2}}{N^{2}} - V(q) \right),
\end{equation}
where
\begin{equation} \label{eq:metric}
 {\gamma}_{\q\q}=-6v_{3}\A\A'^2
\end{equation}
is the target-space metric and
\begin{equation}
{V}=3v_{3}\left(-\A+\lambda ^{2}\A^3\right)
\end{equation}
is the effective potential.\footnote{Note that $\gamma_{\q\q}$ is negative. The kinetic term is therefore of the opposite sign to that of a scalar field. In general, the target space of a minisuperspace model comprising more than one degree of freedom is Lorentzian~\cite{DeWitt:1967yk}. It is well known that due to gauge symmetries, pure Einstein gravity does not yield any propagating ghost. The field $q$ is a background mode that does not correspond to a propagating degree of freedom.  However, due to time-reparametrization invariance, {\it the Hamiltonian is constrained to vanish} and the only allowable solutions are the ones corresponding to zero energy. No large negative energy solution is allowed.} The conjugate momentum of the field $q$ is given by 
\begin{equation}
 \pi _{q} = \gamma _{\q\q}\,\frac{\dot{q}}{N}
\end{equation}
and the classical Hamiltonian satisfies
\begin{equation} \label{eq:H}
\begin{split}
H&=\pi_q \dot q - L\noindent\\
&=  {1\over 2}\, \gamma_{\q\q}\, {\dot q^2\over N}+ NV \noindent\\
& =-N\,{\partial L\over \partial N}\,.
\end{split}
\ee
Since $\partial L/\partial N=0$ is imposed,  the on-shell classical Hamiltonian is constrained to vanish,
\begin{equation} \label{hamiltonian2}
\frac{H}{N}\equiv   \frac{1}{2}\, \gamma ^{\q\q} \pi _{q}^{2} + V   = 0\,.
\end{equation}
This can also be seen as a consequence of time-reparametrization invariance. 


\section{Ambiguities of the canonical-quantization}
\label{quanti}

Let us now apply the canonical approach to quantize the model. The aim is to derive the WDW equation satisfied by the possible wavefunctions of the Universe. We will review the fact that this approach leads to ambiguities in the precise expression of the equation.  

In the canonical approach, the field  $q$ and its conjugate momentum $\pi_q$ are promoted to quantum operators, 
\begin{equation} \label{canmom1}
  q\longrightarrow \widehat{q}\, ,\qquad  \pi_{q} \longrightarrow \widehat{\pi}_q\, , 
\end{equation}
which satisfy the commutation relation
 \begin{equation} \label{canonical1}
\begin{split}
    [\widehat{q}, \widehat{\pi}_{q}] &= i\hbar\,.
\end{split}
\end{equation}
This algebra can be represented by considering  a Hilbert space of wavefunctions  that describe the possible states of the Universe. Since these functions depend on the coordinate $\q$ of the target space $\T$,  the commutation relation is achieved as usual by substituting\footnote{Notice that the inverse minisuperspace metric appearing in the kinetic term of the Hamiltonian may yield powers of the inverse of the position operator $\hat q$. As usual, the domain of the position operator $\hat q$ can be considered to be the larger space of tempered distributions and its eigenfunctions the Dirac $\delta$-function distributions centered at the coordinate positions $\q$ (with eigenvalues $\q$). These eigenfunctions are non-normalizable. Notice that since $a={\cal{A}}(\q)\ge 0$,  we have $(\sign{\A})(\q -\q_*) \ge 0$, where $\q_*=\A^{-1}(0)$ and $\sign{\A}$ is constant since $\A$ is invertible.  The constraint $(\sign{\A})(\q -\q_*) \ge 0$ can be implemented by using a projection operator $\hat \Pi = \Theta((\sign{\A})(\q -\q_*))$, where $\Theta$ is the Heaviside step function. If $(\sign\A)\, \q_* > 0$, the position operator is invertible along the half line $(\sign{\A})(\q -\q_*)\ge 0$. Otherwise, we have $(\sign\A)\,\q_* \le  0$. In this second case, the inverse position operator is singular at $\q=0$ and  is thus defined for $(\sign{\A})(\q -\q_*)\ge 0$, $\q\neq 0$. However, despite the singularity at $\q = 0$, the wavefunction solution of the WDW equation and the corresponding probability density can be well behaving there, resolving this singularity. In \Refe{PTV}, we show that this is indeed the case.}
\begin{equation}
\label{qmomentum}
\widehat q\longrightarrow \q\, , \qquad \widehat{\pi}_q\longrightarrow  -i\hbar \,\frac{\d}{\d \q}\,.
\end{equation}
The constraint of cancellation of the on-shell classical Hamiltonian is promoted at the quantum level  by requiring that the quantum Hamiltonian $\widehat{H}$ annihilates any wavefunction $\psi(\q)$ of the Universe, 
\begin{equation}
\frac{\widehat{H}}{N}\, \psi = 0\,.
\end{equation}
Using \Eqs{hamiltonian2},~(\ref{canmom1}) and~(\ref{qmomentum}), this constraint becomes the WDW equation for $\psi(\q)$.

As is well known, this quantization procedure leads however to ambiguities in the  expression of the WDW equation~\cite{DeWitt:1967yk, Hawking:1985bk}. They result from the existence of various possible choices regarding the ordering of the operators. In practice, two arbitrary complex functions can be introduced in the kinetic term of the classical Hamiltonian in \Eq{hamiltonian2} \cite{PTV}. Denoting them as $v^\q(q)$ and $s(q)$ for reasons that will be clear in the sequel, we have indeed
\begin{equation} \label{eq:ambiguities}
    \gamma ^{\q\q} \pi _{q} \pi _{q} = \frac{ \gamma ^{\q\q} }{v^\q \, s}\, \pi_{q} \, v^\q \, \pi _{q} \, s\, , 
\end{equation}
since every term commutes. However, upon quantization, the right-hand side accounts for a variety of distinct operators, since the commutators of the momentum operators and the ambiguity functions are non-trivial. Note that assuming the WDW equation to be second order in derivatives, no ambiguity function (also) depending on $\pi_q$ can be introduced in the kinetic or potential term. As a result, the WDW equation takes the general form
\begin{equation}\label{wdw}
 \frac{\widehat{H}}{N}\,\psi\equiv -\frac{\hbar ^{2}}{2} \frac{\gamma^{\q\q}}{v^\q\, s}\frac{\d}{\d \q}\bigg(v^\q\,\frac{\d(s\psi)}{\d \q}\bigg)+V\psi =0\,,
\end{equation}
where all quantities are now depending on the target-space coordinate $\q$. Let us now implement a change of variable $\q\equiv \q(\check \q)$. In terms of the target manifold $\cal T$, this is a change of coordinate. Under such a transformation, the wavefunction is a scalar and the ambiguity functions appearing in the WDW equation transform as
\be
s\longrightarrow s\, , \qquad v^\q\longrightarrow v^\q\, {\d \check \q\over\d \q}\equiv v^{\check \q}\, . 
\ee
In other words, $s$ is a scalar and $v^\q$ is a vector, thus motivating the upper index notation anticipated before. We find it convenient to redefine the ambiguity {vector and scalar} as follows:~ 
\begin{equation}
\rho^\q = v^\q s^2 \,,\qquad  \omega = {\gamma^{\q\q}\over v^\q \,s}{\d\over \d \q}\!\left(v^\q {\d s\over \d \q}\right)\equiv {\nabla^\q(v^\q \,\nabla_{\!\q} s)\over v^\q \,s}\,,
\end{equation}
where $\nabla_{\!\q}$ is the covariant derivative on $\cal T$. In terms of these quantities, the WDW equation takes the alternative diffeomorphism-invariant forms
\be\label{wdwq}
\begin{split}
0= \frac{\widehat{H}}{N}\,\psi&\equiv  -\frac{\hbar ^{2}}{2} \!\left[{\gamma^{\q\q}\over \rho^\q} \frac{\d}{\d \q}\! \left(\rho^\q\, \frac{\d\psi}{\d \q}\right) \!+ \omega \psi \right]\!+V\psi\\
& =  -\frac{\hbar ^{2}}{2}\!\left[\gamma^{\q\q}\psi^{\prime\prime}+\gamma^{\q\q}\,{\rho^{\q\prime}\over\rho^\q}\,\psi'+\omega\psi\right]\!+V\psi\esp\\
& =  -\frac{\hbar ^{2}}{2}\!\left[\nabla^2\psi+{\nabla^\q\rho^\q\over\rho^\q}\nabla_{\!\q}\psi+\omega\psi\right]\!+V\psi\, ,\esp
\end{split}
\ee
{where primes now stand for derivatives with respect to the variable $\q$.
Note that the ambiguity vector  $\rho^\q$ is not an artefact that can be gauged away by a change of coordinate in $\T$. Indeed, if we implement the change of variable $\q\equiv\q(\check\q)$ such that $\rho^{\check \q} =1$, the inverse metric $\gamma^{\q\q}$ transforms into an expression of $\gamma^{\check \q\check \q}$ that depends on the initial ambiguity vector~$\rho^\q$. }


\section{Ambiguity of the path-integral quantization}
\label{path}

Quantization of the minisuperspace model of \Sect{classmod} can also be achieved from a path-integral point of view---see \eg \cite{Hartle:1983ai, Halliwell, Halli2, Turok1, Halli-Hartle, PTV}. In this section, we explain that this  again leads to ambiguity \cite{PTV}. 

In this approach, one considers all paths for the lapse function $N$ and field $q$ that are defined on arbitrary intervals of time $[x^0_\i,x^0_\f]$ and satisfying the boundary conditions 
\be
q(x^0_\i)=\q_\i\, , \qquad  q(x^0_\f)=\q\, .
\ee 
Since the WDW equation is linear and of order 2, the Hilbert space of wavefunctions  is two-dimensional and the boundary condition which involves some given $\q_\i$ is enough to specify the state of the Universe \cite{Hartle:1983ai}. A   formal expression of the wavefunction is then given by~\cite{PTV}
\be
\psi(\q)=\int {\D N \over \Vol(\Diff[-N^2])} \int_{q(x^0_\i)=\q_\i}^{q(x^0_\f)=\q}\D q \; e^{{i\over \hbar}\int_{x^0_\i}^{x^0_\f}\d x^0\, L(N,q,\dot q)}\, ,
\label{psidef}
\ee
where $L(N,q,\dot q)$ is the Lagrangian given in \Eq{minisuperspace}. In the exponent, the action is invariant under time reparametrizations. As a result, when distinct metrics $g_{00}\equiv -N^2$ of the base manifold of the $\sigma$-model are related by a diffeomorphism, they are overcounting physically equivalent configurations. For this reason, the path-integral measure $\D N$ is divided by the volume of $\Diff[-N^2]$, the group of diffeomorphisms. Moreover,  the path-integral measure $\D q$ must be invariant under $\Diff[-N^2]$, for the field configurations $q(x^0)$ and $q(\xi(x^0))$ to be truly equivalent for any $\xi\in \Diff[-N^2]$. 

An important thing to note is that  two distinct metrics $g_{00}$ of the base manifold, which is a line segment, cannot in general be transformed into each other by a diffeomorphism. This can be seen by noticing that such a transformation does not  change the proper length~$\ell$ of a line segment,\footnote{Talking about ``duration'' rather than ``length'' may be more appropriate in the present case.}
\be
\ell = \int_{x^0_\i}^{x^0_\f} \d x^0 \,\sqrt{-g_{00}}\, .
\label{di}
\ee  
As a result, the set of metrics $g_{00}$ can be divided into equivalence classes characterized by the value of $\ell\in \R_+$. 
To put it another way, a line segment admits a moduli space of real dimension one parametrized by $\ell$. In practice, varying the modulus of a metric $g_{00}$ amounts to rescaling it, while keeping fixed its domain of definition $[x^0_\i,x^0_\f]$. Hence, denoting $g_{00}[\ell]$ a particular representative of each equivalence class, we may take  
\be
 g_{00}[\ell]=-\ell^2\, , \quad \mbox{defined on}\quad [x^0_\i, x^0_\f]=[0,1]\, .
 \label{l2}
\ee
This corresponds to a gauge choice for which the lapse function is constant, $N\equiv \ell$,  and the time interval is fixed. 
Any metric $g_{00}$ in class $\ell$ can then be obtained by the action of a diffeomorphism $\xi$ on $g_{00}[\ell]$.\footnote{For a line segment, such a transformation $\xi$ is unique up to the discrete isometry that  reverses the orientation of the segment.} 

The above considerations can be used to simplify \Eq{psidef}~\cite{PTV}. Indeed, the path integral over~$N$ can be replaced by an integral over the moduli space and an integral over the orbit of each equivalence class. This has the effect of cancelling the volume of the diffeomorphism group. Following the method of Faddeev and Popov, this implementation of a gauge fixing requires the introduction of a Jacobian $\Delta_{\rm FP}$, which depends only on the class $\ell$. As a result, one obtains an expression
\be
\psi(\q)=\int_0^{+\infty}\!\!\d\ell \; \Delta_{\rm FP}(\ell) \int_{q(0)=\q_\i}^{q(1)=\q}\D q   \; e^{{i\over \hbar}\int_{0}^{1}\d x^0\, L(\ell,q,\dot q)}\, .
\label{ps2}
\ee
In general, the expression of the Faddeev-Popov determinant $\Delta_{\rm FP}$ depends on the topology of the base manifold of the $\sigma$-model. For a line segment, the result turns out to be trivial~\cite{PTV}, \ie independent of the class~$\ell$,\footnote{In other instances, where the base manifold is periodic, one obtains $\Delta_{\rm FP}(\ell)=1/\ell$. }
\be
\label{Delta}
\Delta_{\rm FP}(\ell)=1\, . 
\ee 

Note that in principle, one may determine the ambiguity tensors $\rho^\q$ and $\omega$ by calculating explicitly the path integral in \Eq{ps2} for a certain $\q_\i$, and by imposing that the result is a solution of the  WDW equation.   In practice, though, it is enough to evaluate $\psi(\q)$ at second order in $\hbar$. To see this, let us  consider the formal expansion in $\hbar$,
\be
\label{exp}
\psi(\q)=\exp\!\left[{i\over \hbar}\Big(\F_0(\q)+{\hbar\over i}\,\F_1(\q)-\hbar^2\F_2(\q)+\O(\hbar^3)\Big)\right].
\ee
By computing the path-integral at the semiclassical level, \ie the next-to-leading order in~$\hbar$, one identifies the functions $\F_0(\q)$, $\F_1(\q)$. Then, inserting \Eq{exp} in the second line of \Eq{wdwq}, one determines $\rho^\q$~\cite{PTV}. Since $\omega$ shows up at second order in $\hbar$ in the WDW equation, it can be determined the same way by computing $\F_2(\q)$, \ie the path integral at second order. However, we will not follow this strategy in the present work. On the contrary, we will see in \Sect{resolving} that much less effort is needed to determine the ambiguity tensors of the model we consider. As will be discussed in the last section, this model is representative of a broader class of theories, which are characterized by $\sigma$-models whose target spaces are locally Minkowskian. 

Before doing so, we would like to emphasize an important issue of the path-integral formalism. To present it, let us  implement a field redefinition 
\be
q=Q(\check q)
\ee
in the path integral, where $Q$ is any invertible function. The classical action and thus the integrand in \Eq{ps2} are invariant, while the path-integral measures of the fields $q$ and $\check q$ differ from each other by a Jacobian. Specifically, we have
\be
\D q = \D \check q \;\J\, ,
\ee
where a formal expression of $\J$ is given by 
\be
\label{J}
\J=\prod_{x^0\in[0,1]}\left|{\d q(x^0)\over \d \check q(x^0)}\right|=\prod_{x^0\in[0,1]} \big|Q'(\check q(x^0))\big|\, .
\ee
As a result, we have
\be
\label{psiq}
\begin{split}
\psi(\q)&=\int_0^{+\infty}\!\!\d\ell \int_{\q_\i}^{\q}\D q   \; e^{{i\over \hbar}\int_{0}^{1}\d x^0\, L(\ell,q,\dot q)}\, ,\\
&=\int_0^{+\infty}\d\ell\int_{\check \q_\i}^{\check \q}\D \check q \;\J\, e^{{i\over \hbar}\int_0^1\d x^0\, \check L(\ell,\check q,\dot{\check q})} = \psi(\q(\check\q)) \,,
\end{split}
\ee
in which  we have defined 
\begin{equation} 
\check L(\ell,\check q,\dot{\check q}) = \ell \left({1\over 2}\, \gamma _{\check \q \check \q}(\check q)\, \frac{\dot{\check q}^{2}}{\ell^2} - \check V(\check q) \right), \qquad \where\qquad \check V(\check q)=V\big(Q^{-1}(\check q)\big)\, .
\end{equation}
Moreover, the boundary conditions of the paths $\check q(x^0)$, $x^0\in[0,1]$, are
\be
\check q(1)=\check \q\, , \qquad  \check q(0)=\check \q_\i\, ,
\ee
where we have set 
\be
\begin{split}
\check \q&=Q^{-1}(q(1))=Q^{-1}(\q)\:\equiv \check \q(\q)\, ,  \\
\check \q_\i&= \check \q(\q_\i)\,.
\end{split}
\ee
In fact, in the above equation, the first line defines a change of coordinate $\q=\q(\check \q)$ of the target space $\T$ that we have used in the second line. As a result, $\psi(\q(\check \q))$ is given by the second line in \Eq{psiq} and solves the WDW equation {with the transformed ambiguity vector
\be
\label{change}
\rho^{\check \q}= \rho^\q\, {\d \check \q\over\d \q}
\ee
and the ambiguity scalar $\omega$.} However, had we considered from the outset the field $\check q$ instead of $q$, we would have claimed that the wavefunction is
\be
\check\psi(\check\q)=\int_0^{+\infty}\d\ell\int_{\check \q_\i}^{\check \q}\D \check q \; e^{{i\over \hbar}\int_0^1\d x^0\, \check L(\ell,\check q,\dot {\check q})}\, .
\label{path2}
\ee
Since this expression is identical to the second line of \Eq{psiq} except that the Jacobian is missing, one concludes that $\check\psi(\check \q)$ and $\psi(\q(\check \q))$ are {\it not equal}. Hence, there also exists an ambiguity in the definition of the wavefunction from the path-integral point of view, as there exists a multitude of distinct choices of path-integral measures---such as $\D q$ or $\D\check q$---while the integrand is invariant under field redefinitions. Since $\check\psi(\check \q)$, for a given $\check\q_\i$, is as legitimate as $\psi(\q)$, for the corresponding $\q_\i$, as a wavefunction, it follows that $\check\psi(\check \q)$ is a solution of a WDW equation with ambiguity tensors $\check \rho^{\check \q}$, $\check\omega$ that are  {\it distinct} from $\rho^{\check \q}$, $\omega$. A fundamental question then arises. {\it Is the quantum theory based on $\psi(\q)$ given in Eq.}~(\ref{psiq}) {\it and corresponding ambiguity tensors $\rho^\q$,~$\omega$, equivalent to the quantum theory based on $\check\psi(\check\q)$ given in Eq.}~(\ref{path2}) {\it and corresponding ambiguity tensors $\check\rho^{\check \q}$, $\check\omega$?} 


\section{Inner product of the Hilbert space and probability amplitudes}
\label{inner pro}

To answer the above question, we first have to discuss probability amplitudes. We will complete our arguments in the next section.

Probability amplitudes are based on an inner product we wish to specify. For the Hilbert space of wavefunctions computed for a given choice of path-integral measure $\D q$,\footnote{Or, equivalently, for the Hilbert space of solutions of the WDW equation involving the corresponding ambiguity tensors $\rho^\q$, $\omega$.}  let us define the inner product as 
\begin{equation} 
\label{amp}
\begin{split}
    \langle \psi_1,  \psi_2 \rangle_{\D q} &= \int_\T \d \q \, \sqrt{-\gamma} \, \mu \, \psi_1^* \psi_2\\
    &= \sign(\A')\int^{\A^{-1}(+\infty)}_{\A^{-1}(0)} \d \q \, \sqrt{-\gamma} \, \mu \, \psi_1^* \psi_2\,.
    \end{split}
\end{equation}
In this expression, $\mu(\q)$  is a real positive function that serves as a measure on the Hilbert space of wavefunctions $\psi_1(\q)$, $\psi_2(\q)$, which are defined as in the first line of~\Eq{psiq} for arbitrary initial values $\q_{\i_1}$, $\q_{\i_2}$, respectively. In the second line of \Eq{amp}, the bounds of the integral correspond to the values of $\q$ where the scale factor vanishes or is infinite. Using \Eq{aA}, these values  are $\A^{-1}(0)$ and $\A^{-1}(+\infty)$. Since $\A$ is an invertible function, it is monotonic and therefore the sign of its derivative $\A'$ is fixed. The overall factor $\sign(\A')$ is introduced to make the inner product positive.   

The measure $\mu$ can be determined by requiring the quantum Hamiltonian to be Hermitian~\cite{PTV}. To do this, let us extend the definition~(\ref{amp}) to arbitrary complex functions $f_1(\q)$, $f_2(\q)$. Since they are arbitrary, they don't have to be annihilated by the Hamiltonian. Integrating by parts, one obtains the identity
\be
\label{hermiticity1}
\begin{split}
\left \langle f_1 ,  {\widehat H\over N} \,f_2 \right\rangle = & \left\langle {\widehat H^\dagger\over N}\, f_1 ,  f_2 \right \rangle  -\sign(\A')\,{\hbar^2\over2}\,\times\\
 &~~~~~~~~~~~~\,\left[\rho^\q\left({\gamma^{\q\q}\over \rho^\q}\, \sqrt{-\gamma}\,\mu\, f_1^*\, {\d f_2\over \d\q}-{\d\over \d\q}\!\left({\gamma^{\q\q}\over \rho^\q}\, \sqrt{-\gamma}\,\mu\, f_1^*\right)f_2\right)\right]_{\A^{-1}(0)}^{\A^{-1}(+\infty)}\,,
\end{split}
\ee
where we have defined for any complex function $f$ 
\be\label{wdw+}
 \frac{\widehat{H}^\dag}{N}\,f\equiv -\frac{\hbar ^{2}}{2} {1\over \sqrt{-\gamma}\,\mu}{\d\over \d\q}\!\left(\rho^{\q*} {\d\over \d\q}\Big({\gamma^{\q\q}\over \rho^{\q*}}\, \sqrt{-\gamma}\,\mu\, f \Big)\right)  + \left(V-{\hbar^2\over 2}\,\omega^*\right)f\,.
\ee
Let us suppose for a moment that
\be
\label{HH+}
{\widehat H\over N}\, f={\widehat H^\dag\over N}\, f 
\ee
is satisfied for all $f$. In this case, applying \Eq{hermiticity1} to arbitrary wavefunctions $\psi_1$, $\psi_2$, \ie functions $f_1$, $f_2$ annihilated by $\widehat H$, one concludes that the boundary term in the right-hand side vanishes. Hence, if  \Eq{HH+} is satisfied for all complex function $f$, the Hamiltonian is automatically a Hermitian operator on the Hilbert space of wavefunctions. 

To make all this concrete,  let us define 
\be
\label{polar}
\rho^\q=r^\q\, e^{i\theta}\, , \qquad \where \qquad r^\q\in \R\,, ~~\theta\in\left(-{\pi\over 2},{\pi\over 2}\right] . 
\ee
Since all changes of coordinate $\q\equiv\q(\check\q)$ are real, $r^\q$ transforms as a vector and $\theta$ as a scalar.\footnote{The seemingly more natural choice of taking  $r^\q\ge 0$ and $\theta\in(-\pi,\pi]$ would not lead to tensorial transformations under an invertible change of coordinate $\q=\q(\check\q)$ such that the constant sign of $\d\q/\d\check\q$ is negative.}  
\Eq{HH+}  then leads to 
\be
 2\gamma^{\q\q}\!\left[{r^{\q\prime}\over r^\q}-{(\gamma^{\q\q}\sqrt{-\gamma}\,\mu)'\over \gamma^{\q\q}\sqrt{-\gamma}\,\mu}\right]\!f'=\!\left[-2i \Imp\omega+\gamma^{\q\q}{\dis \Big({\gamma^{\q\q}\over \rho^{\q*}}\sqrt{-\gamma}\, \mu\Big)^{\!\prime\prime}+{\rho^{\q*\prime}\over \rho^{\q*}}\Big({\gamma^{\q\q}\over \rho^{\q*}}\sqrt{-\gamma}\, \mu\Big)^{\!\prime}\over \dis{\gamma^{\q\q}\over \rho^{\q*}}\sqrt{-\gamma}\, \mu}\right]\!f\, .
\label{con}
\ee
If the expression in brackets in the left-hand side does not vanish identically, the equation is separable and $f$ takes a particular form, which contradicts the fact that it is arbitrary. Hence the expression in brackets in the left-hand side cancels out identically, which fixes the measure to be 
\be
\label{mu}
\begin{split}
\mu&={|r^\q|\over -\gamma^{\q\q}\sqrt{-\gamma}}\\
&=\sqrt{-\gamma}\, |\rho^\q|\, , \esps
\end{split}
\ee
up to an irrelevant multiplicative integration constant $C>0$. Indeed, one can always absorb a factor $\sqrt{C}$ in the definition of each wavefunction in \Eq{amp}.  Using the expression for $\mu$,  vanishing of the right-hand side of \Eq{con} also yields 
\be
\label{imo}
\begin{split}
\Imp\omega&={\gamma^{\q\q}\over 2}\!\left(\theta^{\prime\prime}+{r^{\q\prime}\over r^\q}\,\theta'\right)\\
&={1\over 2}\!\left(\nabla^2\theta+{\nabla^\q r^\q\over r^\q}\, \nabla_{\!\q}\theta\right).
\end{split}
\ee
This result means that Hermiticity of the Hamiltonian left us with only three real quan\-ti\-ties---rather than four---to parametrize the ambiguities of the WDW equation, namely $r^\q$, $\theta$, $\Rep\omega$. At this stage, a few remarks are in order:
\begin{itemize}
\item In \Eq{amp}, the volume element $|\d\q|\sqrt{-\gamma}$, the measure $\mu$ and the wavefunctions are invariant under a change of variable $\q\equiv\q(\check\q)$. It is thus consistent to find that the expression~(\ref{mu}) is a scalar under reparametrization of the target space $\T$. 

\item  However, $\mu=\sqrt{-\gamma_{\q\q}}\,|\rho^\q|=\sqrt{-\gamma_{\check\q\check\q}}\,|\rho^{\check\q}|$ depends on the module of the ambiguity vector~$\rho^{\check\q}$. As a result, $\mu$ differs \apriori from the measure $\check\mu=\sqrt{-\gamma_{\check\q\check\q}}\,|\check\rho^{\check \q}|$ associated to the Hilbert space of wavefunctions $\check\psi(\check\q)$, for which the correct ambiguity vector is~$\check\rho^{\check \q}$.\footnote{We say ``{\it a priori}'' because at this stage, there is still the possibility that $\rho^{\check\q}$ and $\check\rho^{\check\q}$ differ only by a phase. However, we will see in the next section that this is not the case.} In other words, both the wavefunctions and the measure on their Hilbert space depend on the choice of path-integral measure---$\D q$ or $\D\check q$. 

\item  Since $\omega$ is invariant under a change of variable $\q\equiv\q(\check\q)$, it is consistent to find an expression for $\Imp\omega$ that is explicitly a scalar, as shown in the second line of~\Eq{imo}. 
\end{itemize}

According to the comments above, it turns out to be relevant to consider the particular combination of Hilbert-space measure and wavefunction 
\be
\label{Pp}
\Psi=\sqrt{\mu}\, \psi\, ,
\ee
in terms of which the inner product of $\psi_1$, $\psi_2$ simplifies to
\be
\label{inner}
\langle \psi_1,  \psi_2 \rangle_{\D q}=\int_\T\d\q\, \sqrt{-\gamma} \, \Psi_1(\q)^*\Psi_2(\q)\,.
\ee
Indeed, using only the definition~(\ref{Pp})---\ie without knowing what the expression of the scalar $\mu$ is---it is straightforward to show that the WDW equation~(\ref{wdwq}) can be written as 
\be
\label{wdw2}
0=\sqrt{\mu}\,{\widehat H\over N}\,\psi\equiv-{\hbar^2\over 2}\big[\nabla^2\Psi+Y^\q\nabla_{\!\q}\Psi+W\Psi\big]\!+V\Psi\, , 
\ee 
where we have defined
\be
\begin{split}
Y^\q&=\gamma^{\q\q}\, {\left(\dis{\rho^\q\over -\gamma^{\q\q}\sqrt{-\gamma}\mu}\right)'\over \dis{\rho^\q\over -\gamma^{\q\q}\sqrt{-\gamma}\mu}}={\nabla^\q\!\left(\dis{\rho^\q\over -\gamma^{\q\q}\sqrt{-\gamma}\mu}\right)\over \dis{\rho^\q\over -\gamma^{\q\q}\sqrt{-\gamma}\mu}}
\end{split}
\ee
and
\be
W=\omega-{\nabla^2\!\sqrt{\mu}\over \sqrt{\mu}}-Y^\q\,{\nabla_{\!\q}\sqrt\mu\over \sqrt{\mu}}\, .
\ee
In the second expression for $Y^\q$, we have replaced the derivative by a covariant derivative, which is justified by the fact that the quantity in parenthesis is a contravariant vector density of weight $-1$. As a result, $Y^\q$ is a vector and $W$ a scalar under reparametrizations of the target space $\T$. Thanks to  the particular form of $\mu$ given  in \Eq{mu}, the expression of $Y^\q$ reduces to  the purely imaginary quantity 
\be
Y^\q=i\nabla^\q\theta\, .
\ee 
Using the expression of $\Imp\omega$  given in \Eq{imo}, the extra potential term of the equation becomes
\be
\label{W}
W=\Rep\omega-{\nabla^2\!\sqrt{\mu}\over \sqrt{\mu}}+{i\over 2}\,\nabla^2\theta\, .
\ee
To sum up, the WDW equation can be recasted in the form
\be
\label{WDWPSI}
-{\hbar^2\over 2}\big[\nabla^2\Psi+i\nabla\theta \nabla\Psi+W\Psi\big]\!+V\Psi=0\, , 
\ee  
where the most general ambiguity is parametrized by the real angle $\theta(\q)$ and the complex quantity $W(\q)$. Although both are scalars under coordinate transformations $\q\equiv\q(\check\q)$ of $\T$, it is more appropriate to think of $\theta$ as an axion of the target space, as only its derivatives matter.\footnote{We stress that in the present context, the ambiguity scalar and axion are really meant for the target space $\T$ of the $\sigma$-model, rather than for (space)time, which is the domain of the $\sigma$-model.}

As in \Eq{path2}, had we considered wavefunctions defined with the path-integral measure $\D\check q$ rather than~$\D q$, we would have ended with a WDW equation for $\check\Psi=\sqrt{\check\mu}\,\check\psi$, involving ambiguities $\nabla^{\check \q}\check\theta$, $\check W$. The key point is that if the ambiguities happen to be invariant under the change of path-integral measure, \ie
\be
\label{am}
\check\theta(\check \q)=\theta(\q(\check\q))+\cst\, , \qquad\check W (\check \q)= W(\q(\check\q)) \, , 
\ee
then the two WDW equations are the same and we can identify their pairwise solutions, 
\be
\check\Psi(\check\q)=\Psi(\q(\check\q))\, .
\ee
Under this scenario, the inner product is invariant, since \Eq{inner} yields 
\be
\label{amp2}
\begin{split}
\langle \psi_1,  \psi_2 \rangle_{\D q}&=\int_\T\d\check \q\, \sqrt{-\gamma_{\check\q\check \q}} \, \Psi_1(\q(\check \q))^*\Psi_2(\q(\check \q))\\
&=\int_\T\d\check \q\, \sqrt{-\gamma_{\check\q\check \q}} \, \check\Psi_1(\check \q)^*\check \Psi_2(\check \q)\\
&=\langle \check\psi_1,  \check\psi_2 \rangle_{\D \check q}\, ,
\end{split}
\ee
and the notion of probability amplitude is universal. In other words, invariance of $\nabla^\q\theta$,~$W$ under the change of path-integral measure is the condition for the quantum theory to be unique, \ie invariant under the ambiguous choice of path-integral measure $\D  q$ in the definition of the wavefunctions. In the next section, we determine $\nabla^\q\theta$, $W$ for any path-integral measure $\D q$ and conclude that the quantum theory is unique. 


\section{Determination of the ambiguities \bm $\rho^\q$, $\omega$ and uniqueness of the quantum theory}
\label{resolving}

In this section, we derive the ambiguity tensors $\rho^\q$, $\omega$ of the WDW equation corresponding  to any choice of path-integral measure in the definition of  the wavefunctions. The result for $\Imp \omega$ turns out to be consistent with~\Eq{imo}. Most importantly, we are able to conclude that all choices of path-integral measure lead to identical quantum predictions. 

To start with, let us consider the probability amplitude for the field $q$ to vary  from an initial value $\q_{\i}$ to a final value $\q$, in a lapse of cosmological time equal to $\ell$. In terms of path integral, it can be expressed as  
\be
\begin{split}
U(\q_{\i},\q,\ell)&=\int_{\q_\i}^{\q}\D q   \; e^{{i\over \hbar}\int_{0}^{1}\d x^0\, L(\ell,q,\dot q)}\\
&=\int_{\q_\i}^{\q}\D q   \; e^{{i\over \hbar}\int_{0}^
\ell\d t\,  \scriptsize \big({1\over 2}\gamma_{\q\q}(q)({\d q\over \d t})^2-V(q) \scriptsize \big)}\, , 
\end{split}
\ee
where, in the second line, we apply a change of gauge to use cosmological time defined above \Eq{friedmann1}. In the following, we derive a differential equation satisfied by this amplitude. 

To this end, we look for  a field redefinition for which the new field $q_0$ has a canonical kinetic term. Writing the relation between $q_0$ and the scale factor as 
\be
a=\A_0(q_0)\, , 
\ee
we see from \Eq{eq:metric} that 
\be
-6v_{3}\A_0(q_0)\A_0'^2(q_0)= -1\, ,
\ee
which leads to 
\be
\label{fr}
\A_0(q_0)=\left({3\over 2}\, {1\over \sqrt{6v_3}}\right)^{2\over 3}|q_0-\q_{0*}|^{2\over3}\, ,\
\ee
where $\q_{0*}$ is an arbitrary integration constant. For the function $\A_0$ to be invertible, we choose the canonical field to vary in the range $[\q_{0*},+\infty]$. It is then straightforward to find the explicit relation between the fields $q$ and $q_0$, since 
\be
\label{Q0}
\begin{split}
q&=\A^{-1}(a)=\A^{-1}(\A_0(q_0))\\
&\equiv Q_0(q_0)\, ,
\end{split}
\ee
where the function $Q_0$ is invertible.  
 In terms of $q_0$, the amplitude takes the form
\be
\label{UU0}
\begin{split}
U(\q_{\i},\q,\ell)&=\int_{\q_{0\i}}^{\q_0}\D q_0 \;\J_0  \; e^{-{i\over \hbar}\int_{0}^
\ell\d t\, \left({1\over 2}({\d q_0\over \d t})^2+V_0(q_0)\!\right)}\\
&\equiv U_0(\q_{0\i},\q_0,\ell)\, ,
\end{split}
\ee
where we have defined new variables
\be
\q_0=Q_0^{-1}(\q)\, , \qquad \q_{0\i}=Q_0^{-1}(\q_\i)\, .
\ee
Moreover, the potential term  is simply
\be
V_0(q_0)=V\big(Q_0(q_0)\big),
\ee
while the Jacobian of the path-integral measure is given by
\be
\J_0=\prod_{t\in[0,\ell]}J\big(q_0(t)\big)\, , \qquad \where\qquad J\big(q_0(t)\big)= \big|Q_0'\big(q_0(t)\big)\big|\, .
\ee

Next, we proceed as in the derivation of the Schr\"odinger equation for a nonrelativistic quantum-mechanical particle moving in a one-dimensional space~\cite{Peskin}, except that we have to keep track of the Jacobian $J$. This requires defining a discretized version of the amplitude $U_0(\q_{0\i},\q_0,\ell)$. This is done by dividing the interval of cosmological time $[0,\ell]$ into $n\ge 1$ slices of duration $\varepsilon$. Moreover, one introduces real numbers $q_{0k}$, $k\in\{0,\dots,n\}$, corresponding to the values of the field $q_0(t)$ at the discrete times $t=k\varepsilon$. Since the path integral is over all trajectories satisfying fixed boundary conditions, we only have to integrate over $q_{0k'}$ for $k'\in\{1,\dots,n-1\}$. Furthermore, as the scale factor $a$ can take values from 0 to $+\infty$, the domain of integration of $q_{0k'}$ is from $\A_0^{-1}(0)=\q_{0*}$ to $\A_0^{-1}(+\infty)=+\infty$. In these notations, the discretized amplitude reads
\be
\label{disc}
\begin{split}
U_0^{(n)}(\q_{0\i},\q_0,\ell)={1\over \N(\varepsilon,\q_{0\i},\q_0)} &\prod_{k'=1}^{n-1} \left(\int_{\q_{0*}}^{+\infty}{\d q_{0k'}\over \N(\varepsilon,\q_{0\i},\q_0)} \, J(q_{0k'})\right)\\
&\!\times e^{-{i\over \hbar}\sum_{k=0}^{n-1}\varepsilon \textstyle[{1\over 2}\textstyle({q_{0(k+1)}-q_{0k}\over \varepsilon}\textstyle)^2+V_0\textstyle({q_{0k}+q_{0(k+1)}\over 2})\textstyle]}  \, ,\\
&\!\!\!\!\!\!\!\!\!\!\!\!\!\!\!\!\!\!\!\!\!\!\!\!\!\!\!\!\!\!\!\!\!\!\!\!\!\!\!\!\!\!\!\!\!\!\!\!\!\!\!\!\!\!\!\!\!\!\!\!\!\!\!\!\!\!\!\!\!\!\!\!\!\where \qquad \varepsilon={\ell\over n} \, , \qquad q_{00}=\q_{0\i}\, , \qquad q_{0n}=\q_0\, .
\end{split}
\ee 
Note that in the discretized version of the path-integral measure, we introduce in the denominator a normalization factor $\N$ for each of the $n$ slices. This is allowed---and will turn out to be mandatory---provided this does not modify the weights of the slices, which are determined by the values of the Jacobians $J(q_{0k'})$. This condition means that $\N$ cannot depend on the integration variables $q_{0k'}$, $k'\in\{1,\dots,n-1\}$, but may depend on the duration~$\varepsilon$ of the slices and the fixed boundary values $\q_{0\i}$, $\q_{0}$. As mentioned below \Eq{psidef}, the path-integral measure $\D q$ must be invariant under time reparametrizations. It is probably possible to show that the discretized version of the path-integral measure implemented in \Eq{disc} gives rise, in the continuum limit, to a measure $\D q$ satisfying this property. However, in the following, we will only assume that this is the case. Indeed, we will present in \Sect{compa} a consistency check of this hypothesis, by showing that our final results are fully compatible with those of~\Refe{PTV}, which uses gauge-invariant measures. 

Let us now consider the amplitude where the last slice is removed. This amplitude takes into account all paths starting from $q_{00}=\q_{0\i}$, and ending at some fixed boundary value~$q_{0(n-1)}$. To be explicit, we have 
\be
\begin{split}
U_0^{(n-1)}(\q_{0\i},q_{0(n-1)},\ell-\epsilon)={1\over \N(\varepsilon,\q_{0\i},\q_0)} &\prod_{k'=1}^{n-2} \left(\int_{\q_{0*}}^{+\infty}{\d q_{0k'}\over \N(\varepsilon,\q_{0\i},\q_0)} \, J(q_{0k'})\right)\\
&\!\!\times e^{-{i\over \hbar}\sum_{k=0}^{n-2}\varepsilon \textstyle[{1\over 2}\textstyle({q_{0(k+1)}-q_{0k}\over \varepsilon}\textstyle)^2+V_0\textstyle({q_{0k}+q_{0(k+1)}\over 2})\textstyle]}  \, ,
\end{split}
\ee 
which is consistent since 
\be
\varepsilon={\ell-\varepsilon\over n-1}\, .
\ee
Let us point out that while removing one slice, we make the choice to retain the path-integral measure used previously.  This means that we have $n-1$ factors $\N$ taken at $(\varepsilon,\q_{0\i},\q_0)$ rather than at $(\varepsilon,\q_{0\i},q_{0(n-1)})$. As a result, we have 
\be
\label{UU}
\begin{split}
U_0^{(n)}(\q_{0\i},\q_0,\ell)= \int_{\q_{0*}}^{+\infty}{\d q_{0(n-1)}\over \N(\varepsilon,\q_{0\i},\q_0)} \, J(q_{0(n-1)})\, e^{-{i\over 2\hbar\varepsilon}({\q_{0}-q_{0(n-1)})^2}}  \, &e^{-{i\over \hbar}\varepsilon V_0{\textstyle(}{q_{0(n-1)}+\q_{0}\over 2}{\textstyle)}}  \\ 
&\!\!\!\!\!\!\!\!\!\!\!\!\!\!\! \times\,   U_0^{(n-1)}(\q_{0\i},q_{0(n-1)},\ell-\epsilon)\, .
\end{split}
\ee
We are going to evaluate the right-hand side of this equation for small $\varepsilon$. 

To achieve this, note that for large $n$, \ie small $\varepsilon$, the first exponential in the integrand produces destructive interference, unless $|\q_0-q_{0(n-1)}|\lesssim \sqrt{2\hbar\varepsilon}$. As a result, we perform a double expansion: The first one for small $\varepsilon$ and the second one for $q_{0(n-1)}$ close to $\q_0$. In practice, it is enough to write
\be
e^{-{i\over \hbar}\varepsilon V_0{\textstyle(}{\q_{0}+q_{0(n-1)}\over 2}{\textstyle)}} =1-{i\over \hbar}\,\varepsilon\left[V_0(\q_0)+{\cal{O}}(q_{0(n-1)}-\q_0)\right]\!+\O(\varepsilon^2)
\ee
as well as
\be
\begin{split}
&J(q_{0(n-1)})\, U_0^{(n-1)}(\q_{0\i},q_{0(n-1)},\ell-\epsilon)=\\
&~~~~~~~~~~~\left[1+(q_{0(n-1)}-\q_0)\, {\partial\over \partial q_{0(n-1)}}+{(q_{0(n-1)}-\q_0)^2\over 2}\, {\partial^2\over \partial q_{0(n-1)}^2}+{\cal{O}}\big((q_{0(n-1)}-\q_0)^3\big)\right]\\
&~~~~~~~~~~~~~~~~~~~~~~~~~~~~~~~~~~~~~~~~~~~~~\times \big[J(q_{0(n-1)})\, U_0^{(n-1)}(\q_{0\i},q_{0(n-1)},\ell-\epsilon)\big]\Big|_{q_{0(n-1)}=\q_0}\, .
\end{split}
\ee
We can now integrate term by term over $q_{0(n-1)}$. Since, $\q_{0*}$ is an arbitrary constant, it is convenient to do the computation in the limit where $\q_{0*}\to -\infty$. In this case, the Gaussian integral formul\ae
\be
\int_{-\infty}^{+\infty}\d q\, e^{-\alpha q^2}=\sqrt{\pi\over \alpha}\, , \qquad \int_{-\infty}^{+\infty}\d q\, q^2\, e^{-\alpha q^2}={1\over 2\alpha}\sqrt{\pi\over \alpha}
\ee 
can be used. 
It is however necessary to implement a regularization scheme that ensures convergence of the integrals. This is done by identifying
\be
\alpha={i\over 2\hbar\varepsilon}(1-i\kappa)\, , \qquad \where \qquad \kappa>0\, , 
\ee
and then taking the limit $\kappa\to 0_+$. By proceeding in this way, one obtains
\be
\label{dif}
\begin{split}
U_0^{(n)}(\q_{0\i},\q_0,\ell)={1\over \N(\varepsilon,\q_{0\i},\q_0)}\,&\sqrt{2\pi\hbar\varepsilon\over i}\,\Big(1-i\,{\varepsilon\over \hbar}\, V_0(\q_0)+{\hbar\varepsilon\over 2i}\,{\partial^2\over \partial q_{0(n-1)}^2}+{\cal{O}}(\varepsilon^2)\Big)\\
&~~~~~~\times\big[J(q_{0(n-1)})\, U_0^{(n-1)}(\q_{0\i},q_{0(n-1)},\ell-\varepsilon)\big]\Big|_{q_{0(n-1)}=\q_0}\, .
\end{split}
\ee
In the limit $n\to +\infty$ for which $\varepsilon\to 0$, the above identity yields
\be
U_0(\q_{0\i},\q_0,\ell)\underset{\varepsilon\to 0}\sim{1\over \N(\varepsilon,\q_{0\i},\q_0)}\, \sqrt{2\pi\hbar\varepsilon\over i}\, J(\q_0)\, U_0(\q_{0\i},\q_0,\ell)\, .
\ee
Therefore, the normalization factor can be any function satisfying this consistency constraint. Choosing
\be
\N(\varepsilon,\q_{0\i},\q_0)=\sqrt{2\pi\hbar\varepsilon\over i}\, J(\q_0)\, ,
\label{N}
\ee
\Eq{dif} can be rewritten as follows:
\be
\begin{split}
&i\hbar\,{U_0^{(n-1)}(\q_{0\i},\q_0,\ell-\varepsilon)-U_0^{(n)}(\q_{0\i},\q_0,\ell)\over -\varepsilon}=\\
&\,~~~~~~~~~~~~~~~~~~~~~~~~~~~{\hbar^2\over 2}\, {1\over J(\q_0)}\, {\partial^2\over \partial q_{0(n-1)}^2}\big[J(q_{0(n-1)})\, U_0^{(n-1)}(\q_{0\i},q_{0(n-1)},\ell-\varepsilon)\big]\Big|_{q_{0(n-1)=\q_0}}\\
&\,~~~~~~~~~~~~~~~~~~~~~~~~~~~+V_0(\q_0)\, U_0^{(n-1)}(\q_{0\i},\q_0,\ell-\varepsilon)+\cal O(\varepsilon)\, .
\end{split}
\ee
By taking the continuum limit $n\to +\infty$, one obtains the partial-differential equation
\be
\begin{split}
&i\hbar\,{\partial \over \partial\ell}\, U_0(\q_{0\i},\q_0,\ell)=\\
&\,~~~~~~~~~~~~\left\{{\hbar^2\over 2}\left[{\partial^2\over \partial\q_0^2}+{2\over J(\q_0)}\, {\d J\over \d\q_0}(\q_0)\, {\partial\over \partial\q_0}+{1\over J(\q_0)}\, {\d^2 J\over \d\q_0^2}(\q_0)\right]+V_0(\q_0)\right\} U_0(\q_{0\i},\q_0,\ell)\, .
\end{split}
\ee

Let us now rewrite this equation in terms of the original variables $\q_\i$ and  $\q$. This is done by using the expression of the Jacobian 
\be
\label{J2}
J(\q_0)=|Q'(\q_0)|=\sign(Q')\, {\d \q\over \d\q_0}\, , 
\ee
where $\sign(Q')$ is a constant sign, since $Q_0$ is invertible and thus monotonic. Indeed, one first obtains
\be
\begin{split}
i\hbar\,{\partial \over \partial\ell}&\, U(\q_{\i},\q,\ell)=\\
&\,\left\{{\hbar^2\over 2}J(\q_0)^2\left[{\partial^2\over \partial\q^2}+{3\over \sign(Q_0')J(\q_0)^2}\, {\d J\over \d\q_0}(\q_0)\, {\partial\over \partial\q}+{1\over J(\q_0)^3}\, {\d^2 J\over \d\q_0^2}(\q_0)\right]+V(\q)\right\} U(\q_{\i},\q,\ell)\, .
\end{split}
\ee
Then, by noticing that
\be
\begin{split}
\gamma_{\q\q}(\q)&=\gamma_{\q_0\q_0}(\q_0)\left({\d\q_0\over \d\q}\right)^2\\
&=-{1\over J(\q_0)^2}\, , 
\end{split}
\ee
the above equation can be written as
\be
\label{Ueq}
\begin{split}
i\hbar\,{\partial \over \partial\ell}\, U(\q_{\i},\q,\ell)&= \left\{ -\frac{\hbar ^{2}}{2}\!\left[\gamma^{\q\q}{\partial^2\over \partial\q^2}+{\gamma^{\q\q}\over P^\q}\,{\d P^{\q}\over\d\q}\,{\partial\over \partial\q}+\Omega \right]\!+V\right\}U(\q_{\i},\q,\ell) \\
&\equiv\widehat\H \,U(\q_{\i},\q,\ell)\, .\esp
\end{split}
\ee
In this expression,  we have defined
\be
\label{PO}
\begin{split}
P^\q&= (-\gamma_{\q\q})^{-{3\over 2}}\, ,  \\
\Omega &=  - {1\over 2\,(\gamma_{\q\q})^2}\,{\d^2\gamma_{\q\q}\over \d\q^2} + {1\over (\gamma_{\q\q})^3}\left({\d\gamma_{\q\q}\over \d\q}\right)^2\, ,\esp
\end{split}
\ee
while the operator $\widehat \H$ is interpreted as a quantum Hamiltonian, in the cosmological-time gauge. 
Notice the similarity between the right-hand side of the differential equation~(\ref{Ueq}) and the second line of~\Eq{wdwq}. From the first line of~\Eq{wdwq}, we see that under an arbitrary  change of coordinate $\q\equiv\q(\check \q)$, the quantity $P^\q$ transforms as a vector, thus justifying its upper index notation, while $\Omega$ is a scalar.\footnote{Let us stress that the right-hand sides of~\Eq{PO} do not say anything about the tensorial nature of the left-hand sides. They are only their actual expressions for the choice of coordinate  $\q$ of the target space $\T$. Moreover, $P^{\check \q}$ is not equal to $(-\gamma_{\check\q\check\q})^{-{3\over 2}}=\check P^{\check \q}$, which is the corresponding quantity found when considering the wavefunctions defined with the path-integral measure $\D\check q$ instead of $\D q$.}

Before going any further, it is interesting to note that the amplitude defined as a path integral can be written in the canonical formalism as  
\be
U(\q_{\i},\q,\ell)=\langle \q|e^{-{i\over \hbar}\widehat \H\, \ell}|\q_\i\rangle \, ,
\label{Ubra}
\ee
where the substitution
\be
\q\longrightarrow \widehat q\, , \qquad   -i\hbar \,\frac{\partial}{\partial \q}\longrightarrow\widehat{\pi}_q
\ee
is understood. To show this, notice first that the right-hand side of~\Eq{Ubra} satisfies the differential equation~(\ref{Ueq}),  and the boundary condition at $\ell=0$
\be
\langle \q|e^{-{i\over \hbar}\widehat \H\,0}|\q_\i\rangle =\langle \q|\q_\i\rangle =\delta(\q-\q_\i)\,.
\label{bc}
\ee
Next, let us consider the discretized version of the amplitude for a single slice of duration $\varepsilon$.  From~\Eq{disc}, we see that it does not involve any integration,
\be
U_0^{(1)}(\q_{0\i},\q_0,\varepsilon)={1\over \N(\varepsilon,\q_{0\i},\q_0)} \,e^{-{i\over \hbar}\varepsilon \textstyle[{1\over 2}\textstyle({\q_{0}-\q_{0\i}\over \varepsilon}\textstyle)^2+V_0\textstyle({\q_{0\i}+\q_{0}\over 2})\textstyle]}  \, .
\ee
Using the expression of the normalization $\N$ given in \Eq{N}, one obtains for small $\varepsilon$
\be
U_0^{(1)}(\q_{0\i},\q_0,\varepsilon)={1\over J(\q_0)}\, {1\over \sigma\sqrt{2\pi}}\, e^{-{(\q_0-\q_{0\i})^2\over 2\sigma^2}}\left(1+{\cal{O}}(\sigma^2)\right) , 
\label{U0s}
\ee
where $\sigma$ is the standard deviation of the Gaussian distribution,
\be
\sigma=\sqrt{\hbar\, \varepsilon\over i}\, .
\ee
Since in the $\varepsilon\to 0$ limit, $U_0^{(1)}(\q_{0\i},\q_0,\varepsilon)$ approaches $U_0(\q_{0\i},\q_0,\varepsilon)$, \Eq{U0s} leads to
\be
U(\q_{\i},\q,0)={\delta(\q_0-\q_{0\i})\over J(\q_0)}= {\delta(\q_\i-\q)\over |{\d\q_0\over \d \q}(\q)|J(\q_0)}= \delta(\q-\q_\i)\, , 
\ee
where we have used~\Eq{J2}. This is the same boundary condition as in~\Eq{bc}, and so one concludes that \Eq{Ubra} is true. 

Since the Faddeev-Popov determinant arising from the gauge fixing of time reparametrizations is trivial, the wavefunction defined in~\Eq{ps2} is related to the amplitude as follows:
\be
\psi(\q)=\int_0^{+\infty}\d\ell \;U(\q_{\i},\q,\ell)\, .
\ee
Therefore, integrating over the length $\ell$ of the time interval both sides of \Eq{Ueq}, one obtains
\be
i\hbar\, \big[U(\q_{\i},\q,\ell)\big]^{\ell=+\infty}_{\ell=0}=  \widehat\H\,\psi\equiv -\frac{\hbar ^{2}}{2}\!\left[\gamma^{\q\q}\psi^{\prime\prime}+\gamma^{\q\q}\,{P^{\q\prime}\over P^\q}\,\psi'+\Omega\psi\right]\!+V\psi\, .
\label{peq}
\ee
However, since the integrand of the amplitude 
\be
U(\q_{\i},\q,\ell)=\int_{\q_\i}^{\q}\D q   \; e^{-{i\over \hbar}\int_{0}^
1\d x^0\, \scriptsize \big({1\over 2}|\gamma_{\q\q}|{\dot q^2\over \ell}+\ell V(q)\scriptsize \big)}
\ee 
is highly oscillating when $\ell\to +\infty$ or $\ell\to 0^+$, making sense of the left-hand side of~\Eq{peq} requires a regularization scheme. To present it, let us use the formal expression of the path-integral,
\be
U(\q_{\i},\q,\ell)=\prod_{x^0\in(0,1)}\int_{-\infty}^{+\infty}\d q(x^0)   \; e^{-{i\over \hbar}{\scriptsize\big(}{1\over 2}|\gamma_{\q\q}|{\dot q^2\over \ell}+\ell V(q)\scriptsize \big)}\, ,
\ee
where $q(0)=\q_\i$, $q(1)=\q$.
In order to take the large $\ell$ limit, we add a small imaginary part to the factor $\ell$ multiplying $V(q(x^0))$, with a suitable $x^0$-dependent sign. To be specific, we define 
\be
\begin{split}
U(\q_{\i},\q,+\infty)&=\lim_{\kappa\to 0^+}\lim_{\ell\to +\infty}\prod_{x^0\in(0,1)}\int_{-\infty}^{+\infty}\d q(x^0)   \; e^{-{i\over \hbar} \scriptsize \big({1\over 2}|\gamma_{\q\q}|{\dot q^2\over \ell}+\ell\scriptsize \big[1-i\kappa\sign(V(q))\scriptsize \big] V(q)\scriptsize \big)}\\
&=\lim_{\kappa\to 0^+}\lim_{\ell\to +\infty}\prod_{x^0\in(0,1)}\int_{-\infty}^{+\infty}\d q(x^0)   \; e^{-{i\over \hbar}{\scriptsize\big(}{1\over 2}|\gamma_{\q\q}|{\dot q^2\over \ell}+\ell V(q)\scriptsize \big)}e^{-{\kappa\over \hbar}\ell |V(q)|}\, .
\end{split}
\ee 
Since for any $x^0\in(0,1)$ the integrand vanishes when $\ell\to +\infty$, we conclude that
\be
U(\q_{\i},\q,+\infty)=0\, .
\ee 
For the limit of small $\ell$, the prescription consists in approaching the origin along the positive imaginary axis. Hence, we define 
\be
U(\q_{\i},\q,0)=\lim_{\kappa\to 0^+}\prod_{x^0\in(0,1)}\int_{-\infty}^{+\infty}\d q(x^0)   \; e^{-{i\over \hbar} \scriptsize \big({1\over 2}|\gamma_{\q\q}|{\dot q^2\over i\kappa}+i\kappa V(q)\scriptsize \big)}\, . 
\ee
Since the integrand vanishes when $\kappa\to 0^+$, we obtain
\be
U(\q_{\i},\q,0)=0\, .
\ee 
As a result, \Eq{peq} is nothing but the WDW equation~(\ref{wdwq}), and thus $P^\q$, $\Omega$ are the ambiguity vector and scalar  $\rho^\q$, $\omega$, while the quantum Hamiltonian $\widehat \H=\widehat H/N$ is constrained to vanish. 

We are now ready to present the main results of the present work. The wavefunctions defined in \Eq{psidef}---or in the first line of \Eq{psiq}---for any choice of path-integral measure $\D q$ satisfy the WDW equation
\be
\label{wdwq2}
\begin{split}
&-\frac{\hbar ^{2}}{2}\!\left[\gamma^{\q\q}\psi^{\prime\prime}+\gamma^{\q\q}\,{\rho^{\q\prime}\over\rho^\q}\,\psi'+\omega\psi\right]\!+V\psi=0\, , \\
\mbox{\ie}\,~\quad &~-\frac{\hbar ^{2}}{2}\!\left[\nabla^2\psi+{\nabla^\q\rho^\q\over\rho^\q}\nabla_{\!\q}\psi+\omega\psi\right]\!+V\psi=0\, ,\esp\\
\where\qquad\rho^\q= &(-\gamma_{\q\q})^{-{3\over 2}}\, ,  \qquad
\omega =  - {1\over 2\,(\gamma_{\q\q})^2}\,{\d^2\gamma_{\q\q}\over \d\q^2} + {1\over (\gamma_{\q\q})^3}\left({\d\gamma_{\q\q}\over \d\q}\right)^2\, .\esps
\end{split}
\ee
From the first line of \Eq{wdwq}, one sees that $\rho^\q$ is a vector and $\omega$ is a scalar under an arbitrary change of coordinate $\q(\check\q)$ in the equation, while keeping the same path-integral measure $\D q$ in the definition of the wavefunctions.
Since the ambiguity vector $\rho^\q$ and scalar~$\omega$ are real, the ambiguity axion is simply
\be
\theta = 0\, , 
\ee
and the expression of $\Imp \omega$ given in \Eq{imo} is trivially satisfied. The Hilbert-space measure appearing in  the inner product~(\ref{amp}) is
\be
\label{mures}
\begin{split}
\mu&=\sqrt{-\gamma}\, |\rho^\q|\, \\
&=(-\gamma_{\q\q})^{-1}\, .
\end{split}
\ee 
In the first line, even if $\rho^\q$ is positive, we keep the absolute value in order for the formula to remain true after implementation of a change of coordinate as in~\Eq{change}. In this case, $\mu$ transforms consistently as a scalar. We remind that the expression of $\mu$ in the second line, as well as those of $\rho^\q$, $\omega$ in the second line of \Eq{wdwq2}, are the actual values of these quantities but do not capture their tensorial properties.
Using the fact that $\nabla^2\sqrt{\mu}={1\over \sqrt{-\gamma}}{\d\over \d\q}\big(\!\sqrt{-\gamma}\, \gamma^{\q\q}{\d\sqrt{\mu}\over \d\q}\big)$, one can see that the ambiguity scalar satisfies the relation
\be
\omega = {\nabla^2\!\sqrt{\mu}\over \sqrt{\mu}}\, ,
\ee
which is explicitly a scalar. Remarkably, the extra potential term $W$ defined in~\Eq{W} turns out to be
\be
W=0\, .
\ee
As explained at the end of the previous section, since $\theta$ and $W$ are independent of the choice of path-integral measure, we conclude that the quantum theory is unique, \ie also independent of this choice. In fact, the WDW equation can be put in the alternative form
\be
\label{final}
-{\hbar^2\over 2} \nabla^2\Psi+V\Psi=0\, ,\qquad \where\qquad \Psi=\sqrt{\mu}\,\psi\, .
\ee  
Therefore, the inner product given in \Eq{inner} does indeed take a universal form, in terms of the wavefunctions $\Psi$. 

Note, though, that these conclusions have been reached assuming that the discretized version of the path-integral measure introduced in \Eq{disc} yields a time-reparametrization invariant measure $\D q$ in the continuum limit. In the next section, we provide a non-trivial consistency check of this assumption.


\section{Comparison with previous works}
\label{compa}

In this section, we compare our results with previous literature. We first show that they are compatible with the derivations presented in \Refe{PTV}. We then explain why we disagree about earlier attempts to derive the exact WDW equation for minisuperspace models. As a consequence, our final answer for the equation associated to the model we consider differs to that found in previous works.

At first glance, the expression of the ambiguity vector $\rho^\q$ given in \Eq{wdwq2} seems in contradiction with that found in~\Refe{PTV}. Let us explain that both results are actually fully consistent. As said below~\Eq{exp}, the ambiguity tensors $\rho^\q$ and $\omega$ can indeed be determined by following a  different route to that followed in the present work. In \Refe{PTV}, the wavefunction~$\psi(\q)$, for any choice of path-integral measure $\D q$, was considered in the particular case where \mbox{$\q_\i=\A^{-1}(0)$}. From~\Eq{aA}, this choice corresponds to a sum over all paths $q(x^0)$, $x^0\in[0,1]$, such that the scale factor $a(x^0)$ satisfies $a(0)=0$.\footnote{Actually, \Refe{PTV} considers the Euclidean version of the path integral, as initially introduced by Hartle and Hawking as the ``no-boundary proposal''~\cite{Hartle:1983ai}. } The wavefunction exhibits distinct behaviors~\cite{Hartle:1983ai, PTV}, depending on the regime in which the variable $\q$ is taken, namely
\be
\begin{split}
&\mbox{the classically forbidden region} \quad 0\le\A(\q)< \lambda^{-1}\, , \\
&\mbox{the classically allowed region} \qquad \lambda^{-1}\le \A(\q) \, .
\end{split}
\ee
In the former case, the value $\A(\q)$ of the scale factor cannot be reached by the classical de Sitter evolution in \Eq{scalefactor}, whereas  in the second, it can. \Refe{PTV} actually computes at the semiclassical level the wavefunction as a path integral, for $\q$ in the core of the classically forbidden region.\footnote{The computations are valid when $\A(\q)$ is not too close from 0 and $\lambda^{-1}$.} For these values of $\q$, the ambiguity vector of~\Refe{PTV}, which we denote by $\rho^\q_{\mbox{\tiny\!\!\cite{PTV}}}$, turns out to be 
\be
\label{rhoB}
\begin{split}
\rho^\q_{\mbox{\tiny\!\!\cite{PTV}}}&=\A^{-{3\over 4}}\,  |\A'|^{-{3\over 2}}\\
&=\left(\!{-{\gamma_{\q\q}\over  6v_3}}\right)^{-{3\over 4}}\, , 
\end{split}
\ee
where \Eq{eq:metric} is used in the second line. Up to an irrelevant multiplicative factor, this is the square root of the result found in the present work. To understand that this apparent discrepancy is not a contradiction, note that for a given field redefinition $a=\A(q)$ in the classical action, the path-integral measure chosen in~\Refe{PTV} may differ from that chosen in the present work. Indeed, in~\Refe{PTV}, the quantum fluctuations of the field $q$ are expanded in an orthonormal basis of functions, and the path-integral measure is defined as the exterior product of the differentials of the coefficients of this expansion. Hence, the latter depends on the choice of orthonormal basis. 

To make the above remarks concrete, let us consider the wavefunction defined as in the right-hand side of \Eqs{ps2}, with $\Delta_{\rm FP}=1$, but with the path-integral measure $\D q$ replaced by $\D\tilde q$, which is that associated to a redefined field~$\tilde q$. In this case, the wavefunction is a solution of a WDW equation with ambiguity vector 
\be
\tilde \rho^{\tilde \q}=(-\gamma_{\tilde \q\tilde \q})^{-{3\over 2}}\, ,
\ee 
as follows from \Eq{wdwq2}. Rewriting the equation in terms of the variable $\q$, the ambiguity vector becomes
\be
\tilde \rho^\q=\tilde \rho^{\tilde \q}\, {\d \q\over \d\tilde \q}\, .
\ee
Let us now choose the field $\tilde\q$ so that the associated target-space metric satisfies
\be
\begin{split}
-\gamma_{\tilde \q\tilde \q}&=(-\gamma_{\q\q})^\nu\\
&=-\gamma_{\q\q}\, \Big({\d \q\over \d\tilde \q}\Big)^2\, , 
\end{split}
\ee
for some real coefficient $\nu$. We thus have
\be
{\d\tilde \q\over \d\q}=\pm (-\gamma_{\q\q})^{1-\nu\over 2}\, , 
\ee
where the overall sign is constant, for the change of variable $\tilde\q\equiv \tilde \q(\q)$ to be monotonic. This yields
\be
\tilde \rho^\q=\pm (-\gamma_{\q\q})^{-\nu-{1\over 2}}\, .
\ee
Choosing $\nu=1/4$, we find that 
\be
\label{rhos}
\rho^\q_{\mbox{\tiny\!\!\cite{PTV}}}=\pm  (6v_3)^{3\over 4} \tilde \rho^\q\, .
\ee
This shows that for any choice of fields $q$, denoting $\D_{\mbox{\tiny\!\!\cite{PTV}}} q$ the path-integral measure used in \Refe{PTV}, we have
\be
\D_{\mbox{\tiny\!\!\cite{PTV}}} q=\D \tilde q\, , \qquad \where\qquad \tilde\q=\tilde \q_* \pm \int_{\A^{-1}(0)}^{\q}\d u \, (-\gamma_{\q\q}(u))^{3\over 8}\, , 
\ee
for any choice of real constant $\tilde \q_*$ and sign $\pm$.   Since $\D_{\mbox{\tiny\!\!\cite{PTV}}}q$ is invariant under time reparametrizations~\cite{PTV}, the above identification provides a consistency check of the assumption made below \Eq{disc}, namely that this is also the case for  the path-integral measures used in the present work. 

Next, we would like to raise issues in previous literature. The first concerns the gauge fixing of time reparametrizations in the path integral. In many works, such as \Refs{Turok1, Halli-Hartle}, the wavefunction is initially considered in the form given in \Eqs{ps2},~(\ref{Delta}). As in \Refs{Halliwell,Halli2}, the authors  then implement the following reparametrization of time and field redefinition 
\be
\label{ex}
\ell \,\d x^0 = {\ell\over a(x^0)}\,\d u\, , \qquad a=\A(q)=\sqrt{q}\, .
\ee
Choosing for instance $u(0)=0$, the minisuperspace action  becomes
\be
S={3v_3}\int_0^{u(1)}\d u \left[- {1\over 4\ell} \Big({\d q\over \d u}\Big)^2+\ell\, (1-\lambda q) \right] , \quad ~~\where\quad~~ u(1) = \int_0^1\d x^0\sqrt{q(x^0)}\, .
\label{acu}
\ee
Notice that the upper bound of the integral, $u(1)$, depends on the path of the field $q$. However, many authors \cite{Turok1, Halli-Hartle} replace $u(1)$ by a fixed value equal to 1. As a result, the path integral on the field~$q$ becomes Gaussian and can be computed exactly. However, changing $u(1)$ by~1 is not allowed, as this modifies the classical action and thus the model. 

Moreover, in \Refe{Halliwell},  the attempt in deriving the WDW equation leads to a wavefunction that depends on the gauge of time reparametrizations. This may sound weird, as we are more familiar with computations of path integrals that give an answer independent of the gauge chosen for each local symmetry. In fact, the issue is related to the remark in the previous paragraph. When the gauge choice involves the path of the scale factor and $u(1)$ is replaced by a fixed value, the quantity $\ell$ does not  parametrize anymore the moduli space of the metric~$g_{00}$ of the line-segment of time. Hence, the integral over $\ell$ is not a sum over all inequivalent classes of metrics, implying the final expression for the wavefunction to be incorrect. In fact, each time the gauge of time reparametrizations involves the path while $u(1)$ is replaced by a given value, the result is  the correct wavefunction but for a different model, as stressed before. 

In addition,  it is  postulated in \Refe{Halliwell} that for any minisuperspace model involving $m$ degrees of freedom, the WDW equation should be invariant under field redefinitions of the $m$ fields.\footnote{Note that in our work, we do not postulate such a thing. On the contrary, we show that this is the case at least in a model with $m=1$ degree of freedom.}  It is then concluded that the equation is restricted to take the form 
\be
\label{wdwH}
-\frac{\hbar ^{2}}{2}\!\left[\nabla^2\psi+\xi {\cal R}\psi\right]\!+V\psi=0\, , 
\ee
where ${\cal R}$ is the Ricci scalar of the $m$-dimensional target space $\T$, and $\xi$ is a constant. Notice however that this form is not the most general one involving only terms with at most two derivatives.  Among many possibilities, a term $\xi' {\cal R}V\psi$ for any constant $\xi'$, or a term involving any function of the potential, \ie $f(V)\psi$ rather than $V\psi$, are allowed. Moreover, \Refe{Halliwell} imposes that \Eq{wdwH} and its solutions are proportional to their counterparts, when one changes the lapse function $N$ to $\tilde  N=\gamma^2\tilde N$, where $\gamma$ is an arbitrary function of the $m$ degrees of freedom.\footnote{For a more complete discussion concerning this symmetry and invariance under conformal rescalings of the minisuperspace metric in the higher dimensional cases, see \Refe{Anderson:2009yc}. Such a symmetry may play a role in fixing ordering ambiguities in the higher-dimensional cases.} The idea is that the resulting equation and wavefunctions should give a physical description equivalent to the initial one. It is then claimed that this constraint determines what the value of $\xi$ is.  However, nowhere in the derivation, the bounds of the integrals which define the actions are specified and the issues raised in the two previous paragraphs are ubiquitous. Moreover, as explained in our work, to conclude that \apriori different Hilbert spaces of wavefunctions are equivalent, their inner products must be equal. However, inner products are not discussed in \Refe{Halliwell}. 

In order to compare our results to those of \Refe{Halliwell} in a concrete way, let us write the WDW equation~(\ref{final}) for the wavefunction $\Psi(\q)$, when the variable $\q$ is associated to the field $q$ defined by the relation  $a=\A(q)\equiv \sqrt{q}$:
\begin{equation}
\label{exeq}
    \Psi'' + {1\over 4q}\,\Psi' + {9v_3^2\over \hbar^2} \,(\lambda^2 q -1 )\Psi = 0\, .
\end{equation}
In the particular case where the cosmological term vanishes, $\lambda =0$, an analytic expression of the solutions is given by
\be
\label{solex}
    \Psi(\q) = c_1 \,\q^{3\over 8} J_{3\over 8} \!\left({3v_3\over i \hbar}\,\q \right) + \,c_2 \,\q^{3\over 8} Y_{3\over 8} \!\left({3v_3\over i\hbar}\,\q \right),
\ee
where $J_{3\over 8}$ and $Y_{3\over 8}$ are respectively Bessel functions of the first and second kind, and $c_1$, $c_2$ are arbitrary complex integration constants. On the contrary, with the conventions given in \Eq{ex}, it is claimed in  \Refe{Halliwell} that the minisuperspace model is  Gaussian and that the WDW equation, which is copied in  \Eq{wdwH},  reduces to  
\be
4 \hbar^2 \psi''+(\lambda^2 \q-1)\psi=0\, .
\ee
Its general solution can be explicitly written in terms of Airy functions. In the particular case where $\lambda=0$, the solutions are simply
\be
\psi(\q)=c_1\, e^{\q\over 2\hbar}+c_2\,e^{-{\q\over 2\hbar}}\, , 
\ee
where $c_1$, $c_2$ are complex integration constants. Of course, this disagrees with our results given in \Eqs{exeq} and~(\ref{solex}). However, we would like to stress  that all the problems mentioned above do not disqualify many of the important and seminal ideas of earlier works, on which our results are also based. 


\section{Discussion and conclusion}
\label{conclu}

For closed Universes, the minisuperspace model analyzed in this work is the simplest one, as it involves only the dynamics of a single degree of freedom~$q$. In both the canonical and the path-integral formulations of the quantum theory, there are ambiguities in the precise form of the WDW equation. In the canonical formalism, the ambiguities arise from operator-ordering ambiguities in the expression of the quantum Hamiltonian. In the path-integral point of view, they arise from the inequivalent choices for the path-integral measure linked to field redefinitions of the minisuperspace degree of freedom. In this work, we have resolved the ambiguities by determining the WDW equations for the path-integral wavefunctions at the exact level in $\hbar$. Utilizing the inner-product measure that implements the Hermiticity condition for the quantum Hamiltonian, we have established that the various quantum prescriptions are in fact equivalent, at the exact level in $\hbar$, since they yield the same probability amplitudes and physical observables. No special boundary conditions need to be imposed to select among the quantum prescriptions. The results in this work establish the universality of the prescriptions to all orders in $\hbar$, generalizing results at the semiclassical level. They are reminiscent of a similar property of quantum field theory in that the S-matrix is invariant under field redefinitions despite the many choices for the path-integral measure.   

The reader may wonder about the generality of our approach to resolve the ambiguities of the WDW equation associated to more involved systems. For models involving $m>1$ degrees of freedom $q^i$, $i\in\{1,\dots,m\}$, the target space~$\T$ of the $\sigma$-model is a Lorentzian manifold of dimension~$m$. It turns out that our method applies, provided $\T$ is locally the Minkowski space~$\R^{1,m-1}$. Indeed, in this case only, there exists a field redefinition $q^i=Q_0^i(q_0^1,\dots,q_0^m)$ such that  the metric $\gamma_{\q_0^i\q_0^j}$ of the $\sigma$-model has only constant components. The reason why a constant metric is required is that otherwise, when it is taken at the penultimate slice of the lapse of cosmological time~$\ell$, its expansion around the boundary value $(\q_0^1,\dots,\q_0^m)$ at the last slice is non-trivial. As a result, when evaluating at small $\varepsilon$ the analogue of the right-hand side of \Eq{UU}, one has to integrate a series of terms involving arbitrary positive powers of $1/\varepsilon$. However, Gaussian integrations term by term are not mathematically legitimate and it is not clear how to proceed to obtain a practical result. In these more involved models, it would be interesting to obtain the analogue of \Eq{WDWPSI}, accounting for all ambiguity functions, and to determine the exact form of the analogue of $W$. In particular, it would be worth investigating whether $W\propto \cal R$, where $\cal R$ is the scalar curvature on the minisuperspace manifold, as advocated in \Refs{DeWitt:1967yk,Halliwell}.


\section*{Acknowledgements}
N.T. thanks the Ecole Polytechnique for its hospitality, while H.P. thanks the University of Cyprus for its hospitality. This work is partially supported by the Cyprus Research and Innovation Foundation grant EXCELLENCE/0421/0362.





\begin{thebibliography}{99}

\bibitem{DeWitt:1967yk}
B.~S.~DeWitt,
``Quantum theory of gravity. 1. The canonical theory,''
Phys. Rev. \textbf{160} (1967), 1113-1148
doi:10.1103/PhysRev.160.1113 


\bibitem{Vilenkin:1982de}
A.~Vilenkin,
``Creation of Universes from nothing,''
Phys. Lett. B \textbf{117} (1982), 25-28
doi:10.1016/0370-2693(82)90866-8

\bibitem{Vilenkin:1983xq}
A.~Vilenkin,
``The birth of inflationary Universes,''
Phys. Rev. D \textbf{27} (1983), 2848
doi:10.1103/PhysRevD.27.2848

\bibitem{Vilenkin:1984wp}
A.~Vilenkin,
``Quantum creation of Universes,''
Phys. Rev. D \textbf{30} (1984), 509-511
doi:10.1103/PhysRevD.30.509

\bibitem{Vilenkin:1987kf}
A.~Vilenkin,
``Quantum cosmology and the initial state of the Universe,''
Phys. Rev. D \textbf{37} (1988), 888
doi:10.1103/PhysRevD.37.888

\bibitem{Hartle:1983ai}
J.~B.~Hartle and S.~W.~Hawking,
``Wave function of the Universe,''
Phys. Rev. D \textbf{28} (1983), 2960-2975
doi:10.1103/PhysRevD.28.2960

\bibitem{Hawking:1984hk}
S.~W.~Hawking,
``The cosmological constant is probably zero,''
Phys. Lett. B \textbf{134} (1984), 403
doi:10.1016/0370-2693(84)91370-4

\bibitem{Halliwell:1984eu}
J.~J.~Halliwell and S.~W.~Hawking,
``The origin of structure in the Universe,''
Phys. Rev. D \textbf{31} (1985), 1777
doi:10.1103/PhysRevD.31.1777

\bibitem{Hawking:1985bk}
S.~W.~Hawking and D.~N.~Page,
``Operator ordering and the flatness of the Universe,''
Nucl. Phys. B \textbf{264} (1986), 185-196
doi:10.1016/0550-3213(86)90478-5

\bibitem{Halliwell}
J.~J.~Halliwell,
``Derivation of the Wheeler-De Witt equation from a path integral for minisuperspace models,''
Phys. Rev. D \textbf{38} (1988), 2468. 
doi:10.1103/PhysRevD.38.2468


\bibitem{Halli2}
J.~J.~Halliwell and J.~Louko,
``Steepest descent contours in the path integral approach to quantum cosmology. 1. The de Sitter minisuperspace model,''
Phys. Rev. D \textbf{39} (1989), 2206.
doi:10.1103/PhysRevD.39.2206

\bibitem{Hartle:2008ng}
J.~B.~Hartle, S.~W.~Hawking and T.~Hertog,
``The classical Universes of the no-boundary quantum state,''
Phys. Rev. D \textbf{77} (2008), 123537
doi:10.1103/PhysRevD.77.123537
[arXiv:0803.1663 [hep-th]]

\bibitem{Turok1}
J.~Feldbrugge, J.~L.~Lehners and N.~Turok,
``Lorentzian quantum cosmology,''
Phys. Rev. D \textbf{95} (2017) no.10, 103508
doi:10.1103/PhysRevD.95.103508
[arXiv:1703.02076 [hep-th]]

\bibitem{Halli-Hartle}
J.~Diaz Dorronsoro, J.~J.~Halliwell, J.~B.~Hartle, T.~Hertog and O.~Janssen,
``The real no-boundary wave function in Lorentzian quantum cosmology,''
Phys. Rev. D \textbf{96} (2017) no.4, 043505
doi:10.1103/PhysRevD.96.043505
[arXiv:1705.05340 [gr-qc]]

\bibitem{Linde:1990flp}
A.~D.~Linde,
``Particle physics and inflationary cosmology,''
Contemp. Concepts Phys. \textbf{5} (1990), 1-362
[arXiv:hep-th/0503203 [hep-th]]

\bibitem{Nelson:2008vz}
W.~Nelson and M.~Sakellariadou,
``Unique factor ordering in the continuum limit of LQC,''
Phys. Rev. D \textbf{78} (2008), 024006
doi:10.1103/PhysRevD.78.024006
[arXiv:0806.0595 [gr-qc]]

\bibitem{He:2015wla}
D.~He, D.~Gao and Q.~y.~Cai,
``Dynamical interpretation of the wavefunction of the Universe,''
Phys. Lett. B \textbf{748} (2015), 361-365
doi:10.1016/j.physletb.2015.07.029
[arXiv:1507.06727 [gr-qc]]


\bibitem{PTV}
H.~Partouche, N.~Toumbas and B.~de Vaulchier,
``Wavefunction of the Universe: Reparametrization invariance and field redefinitions of the minisuperspace path integral,''
Nucl. Phys. B \textbf{973} (2021), 115600
doi:10.1016/j.nuclphysb.2021.115600
[arXiv:2103.15168 [hep-th]]

\bibitem{Partouche:2022kfi}
H.~Partouche, N.~Toumbas and B.~de Vaulchier,
``Wavefunction of the Universe: Diffeomeorphism invariance and field redefinitions,''
PoS \textbf{CORFU2021} (2022), 159
doi:10.22323/1.406.0159

\bibitem{Partouche:2021lyb}
H.~Partouche, N.~Toumbas and B.~de Vaulchier,
``Gauge fixing and field redefinitions of the Hartle-Hawking wavefunction path integral,''
doi:10.31526/ACP.BSM-2021.34
[arXiv:2105.04818 [hep-th]]

\bibitem{Kehagias:2021wwr}
A.~Kehagias, H.~Partouche and N.~Toumbas,
``Probability distribution for the quantum Universe,''
JHEP \textbf{12} (2021), 165
doi:10.1007/JHEP12(2021)165
[arXiv:2110.03050 [hep-th]]

\bibitem{Peskin}
M. E. Peskin and D. V. Schroeder, ``An introduction to quantum field theory,'' Perseus Book Publishing (1995).

\bibitem{Henneaux:1992ig}
M.~Henneaux and C.~Teitelboim,
``Quantization of gauge systems,'' Princeton University
Press (1992). 

\bibitem{Anderson:2009yc}
E.~Anderson,
``Relational motivation for conformal operator ordering in quantum cosmology,''
Class. Quant. Grav. \textbf{27} (2010), 045002
doi:10.1088/0264-9381/27/4/045002
[arXiv:0905.3357 [gr-qc]].

\end{thebibliography}
\end{document}